\documentclass[twocolumn,prlprintnumbers,showpacs,prl,floatfix,superscriptaddress,amsmath,amssymb]{revtex4}
\usepackage{epsfig}
\usepackage{subfigure}
\usepackage{amsmath}
\usepackage{color}
\usepackage{amssymb}
\usepackage{setspace}
\usepackage{graphicx}
\usepackage{dcolumn}
\usepackage{bm}
\usepackage{times}
\input epsf

\begin{document}

\author{Daniel B. Larremore}
\email{larremor@colorado.edu}
\affiliation{Department of Applied Mathematics, University of Colorado, Boulder, CO 80309, USA}

\author{Woodrow L. Shew}
\affiliation{National Institutes of Health, National Institute of Mental Health, Bethesda, MD 20892, USA}

\author{Juan G. Restrepo}
\affiliation{Department of Applied Mathematics, University of Colorado, Boulder, CO 80309, USA}

\date{\today}

\begin{abstract}
The collective dynamics of a network of coupled excitable systems in response to an external stimulus depends on the topology of the connections in the network. Here we develop a general theoretical approach to study the effects of network topology on dynamic range, which quantifies the range of stimulus intensities resulting in distinguishable network responses. We find that the largest eigenvalue of the weighted network adjacency matrix governs the network dynamic range. Specifically, a largest eigenvalue equal to one corresponds to a critical regime with maximum dynamic range. We gain deeper insight on the effects of network topology using a nonlinear analysis in terms of additional spectral properties of the adjacency matrix. We find that homogeneous networks can reach a higher dynamic range than those with heterogeneous topology. Our analysis, confirmed by numerical simulations, generalizes previous studies in terms of the largest eigenvalue of the adjacency matrix.
\end{abstract}


\title{Predicting criticality and dynamic range in complex networks: effects of topology}

\pacs{??}

\maketitle

Numerous natural \cite{dendriticTrees,kinouchiCopelli} and social
\cite{epidemics} systems are accurately described as networks of
interacting excitable nodes. The collective dynamics of such
excitable networks often defy naive expectations based on the
dynamics of the single nodes which comprise the network. For
example, the collective response of a neural network can encode
sensory stimuli which span more than 10 orders of magnitude in
intensity, while the response of a single neuron (node) typically
encodes a much smaller range of stimulus intensities. More
generally, the range of stimuli over which a network's response varies significantly is quantified by {\it
dynamic range} and is a fundamental property,
whether the network is comprised of people, cell phones, genes, or neurons.
In neural networks, recent experiments \cite{woody} suggest that
dynamic range is maximized in a critical regime in which
neuronal avalanches \cite{beggsPlenz} occur, confirming earlier
theoretical predictions \cite{kinouchiCopelli}. It has been argued
\cite{kinouchiCopelli,woody} that this critical regime occurs when
the effective mean degree of the network is one, i.e. the
expected number of excited nodes produced by one excited node is
one. However, this criterion is invalid for networks with
broad degree distributions \cite{copelliCampos,wu}. A general
understanding of how dynamic range and criticality depend on
network structure remains lacking.  In this Letter, we present a
unified theoretical treatment of stimulus-response relationships
in excitable networks, which holds for diverse networks including
those with random, scale free, degree-correlated, and assortative topologies.

As a tractable model of an excitable network, here we consider the
Kinouchi-Copelli model \cite{kinouchiCopelli}, which consists of
$N$ coupled excitable nodes. Each node $i$ can be in one of $m$
states $x_{i}$. The state $x_{i}=0$ is the resting state,
$x_{i}=1$ is the excited state, and there may be additional
refractory states $x_{i}=2,3,...,m-1$. At discrete times
$t=0,1,...$ the states of the nodes $x_{i}^{t}$ are updated as
follows: (i) If node $i$ is in the resting state, $x_{i}^{t}=0$,
it can be excited by another excited node $j$, $x_{j}^{t}=1$, with
probability $A_{ij}$, or independently by an external process with
probability $\eta$. The network topology and strength of
interactions between the nodes is described by the connectivity
matrix $A = \{ A_{ij} \}$.  In this model, $\eta$ is considered
the stimulus strength.  (ii) The nodes that are excited or in a
refractory state, $x_{i}^{t} \geq 1$, will deterministically make
a transition to the next refractory state if one is available, or
otherwise return to the resting state (i.e. $x_{i}^{t+1} =
x_{i}^{t} + 1$ if $1 \leq x_{i}^{t} < m-1$, and $x_{i}^{t+1} = 0$
if $x_{i}^{t} = m-1$).

An important property of excitable networks is the dynamic range,
which is defined as the range of stimuli that is distinguishable
based on the system's response $F$. Following \cite{kinouchiCopelli}, we quantify the network
response with the average activity $F = \langle f \rangle_{t}$
where $\langle \cdot \rangle_{t}$ denotes an average over time and
$f^{t}$ is the fraction of excited nodes at time $t$. To calculate a system's
dynamic range, we first determine a lower stimulus threshold
$\eta_{low}$ below which the change in the response is negligible, and 
an upper stimulus threshold $\eta_{high}$ above
which the response saturates. Dynamic range ($\Delta$), measured
in decibels, is defined as $\Delta = 10
\log_{10}{\eta_{high}/\eta_{low}}$.  To analyze
the dynamics of this system, we denote the probability that a
given node $i$ is excited at time $t$ by $p_{i}^{t}$. For
simplicity, we will consider from now on only two states, resting
and excited (m=2) \cite{foot}. Then, the update equation for
$p_{i}^{t}$ is
\begin{equation}\label{meanfieldEq}
\addtolength{\belowdisplayskip}{-0.2cm}
\addtolength{\abovedisplayskip}{-0.2cm}
p_{i}^{t+1} = (1 -p_{i}^{t})\left( \eta + (1-\eta) \left [ 1- \prod_{j}^{N} (1 - p_{j}^{t} A_{ij}) \right ] \right)
\end{equation}
which can be obtained by noting that $1 - p_{i}^{t}$ is the
probability that node $i$ is resting at time $t$, and the term in
large parentheses is the probability that it makes a transition to
the excited state. We note that, in writing this probability, we
treat the events of neighbors of node $i$ being excited at time
$t$ as statistically independent.  As noted before
\cite{epidemics,locallyTreeLike,ottPomerance,genes}, this
approximation yields good results even when the network has a
non-negligible amount of short loops.

In Ref. \cite{kinouchiCopelli}, the response $F$ was theoretically
analyzed as a function of the external stimulation probability
$\eta$ using a mean-field approximation in which connection
strengths were considered uniform, $A_{ij}=\sigma/N$ for all $i,
j$. It was shown that at the critical value $\sigma=1$, the
network response $F$ changes its qualitative behavior. In
particular, $\displaystyle \lim_{\eta \to 0} F = 0$ if $\sigma <
1$ and $\displaystyle \lim_{\eta \to 0} F > 0$ if $\sigma > 1$. In
addition, the dynamic range of the network was found to be
maximized at $\sigma=1$.  The parameter $\sigma$ is defined in Refs.~\cite{kinouchiCopelli,woody} as an average branching ratio, written here as $\sigma = \frac{1}{N}\sum_{i,j} A_{ij}
= \langle d^{in} \rangle = \langle d^{out} \rangle$, where
$d^{in}_{i} = \sum_{j} A_{ij}$ and $d^{out}_{i} = \sum_{j} A_{ji}$
are the in- and out-degrees of node $i$, respectively, and
$\langle \cdot \rangle$ is an average over nodes.  For the network
topology studied by Ref. \cite{kinouchiCopelli} $\sigma=1$ marks
the critical regime in which the expected number of excited nodes
is equal in consecutive timesteps.  Such critical branching
processes result in avalanches of excitation with power-law
distributed sizes. Cascades of neural activity with such power-law
size distributions have been observed in brain tissue cultures
\cite{woody}, awake monkeys \cite{awakeMonkeys}, and anesthetized
rats \cite{anesthetizedRats}.  While $\sigma=1$ successfully
predicts the critical regime for  Erd\H{o}s-R\'{e}nyi random
networks \cite{kinouchiCopelli}, this prediction fails in networks
with a more heterogeneous degree distribution
\cite{wu,copelliCampos}. Perhaps more importantly, previous
theoretical analyses \cite{kinouchiCopelli,wu,copelliCampos} do
not account for features that are commonly found in real networks,
such as community structure, correlation between in- and
out-degree of a given node, or correlation between the degree of
two nodes at the ends of a given edge \cite{newmanAssortativity}.
Here, we will generalize the mean-field criterion $\sigma=1$ to
account for complex network topologies.

To begin, we note that $\displaystyle \lim_{\eta \to 0} F = 0$ corresponds to the
fixed point $\vec{p} = 0$ of Eq. (\ref{meanfieldEq}) with $\eta =
0$. To examine the linear stability of this fixed point, we set
$\eta = 0$ and linearize around $p_{i}^{t}=0$, assuming
$p_{i}^{t}$ to be small, obtaining $p_{i}^{t+1}
=\sum_{j}^{N}p_{j}^{t} A_{ij}$. Assuming 
$p_i^{t}=u_{i} \lambda^{t}$ yields
\begin{equation}\label{firstOrderEq}
\addtolength{\belowdisplayskip}{-0.4cm}
\addtolength{\abovedisplayskip}{-0.4cm}
\lambda u_{i} =\sum_{j}^{N}u_{j} A_{ij}.
\end{equation}
Thus, the stability of the solution $\vec{p}=0$ is governed by the largest eigenvalue of the network adjacency matrix, $\lambda$, with $\lambda < 1$ being stable and $\lambda > 1$ being unstable.  Therefore, the critical state described in previous literature, occurring at various values of $\langle d \rangle$, should universally occur at $\lambda = 1$.  Importantly, since $A_{ij} \geq 0$, the Perron-Frobenius theorem guarantees that $\lambda$ is real and positive \cite{PFTheorem}. Other previous studies in \textit{random} networks have also investigated spectral properties of $A$ to gain insight on the stability of dynamics in neural networks \cite{robinsonGray} and have shown how $\lambda$ could be changed by modifying the distribution of synapse strengths \cite{rajanAbbott}. An important implication of Eq.~(\ref{firstOrderEq}) is that, when $p$ and $\eta$ are small enough, $p$ should be almost proportional to the right eigenvector $u$ corresponding to $\lambda$, so we write $p_{i}=Cu_{i} + \epsilon_{i}$, where C is a proportionality constant and the $\epsilon_{i}$ error term captures the deviation of actual system behavior from the linear analysis.
To first order, the constant $C$ is related to the network response $F$ since, neglecting $\epsilon$, we have
\begin{equation}\label{CtoFEq}
\addtolength{\belowdisplayskip}{-0.4cm}
\addtolength{\abovedisplayskip}{-0.2cm}
F = \langle f \rangle_{t} = \frac{1}{N} \sum_{i} p_{i} \approx \frac{1}{N} \sum_{i} C u_{i} = C \langle u \rangle.
\end{equation}

The linear analysis allowed us to identify $\lambda=1$ as the point at which the network response becomes non-zero as $\eta \to 0$. In what follows, we use a weakly nonlinear analysis to obtain approximations to the response $F(\eta)$ when $\eta$ is small. As we will show, these approximations depend only on a few spectral properties of $A$. Assuming $A_{ij} p_{j} \ll 1$ (which is valid near the critical regime if each node has many incoming connections), we approximate the product term of Eq.~(\ref{meanfieldEq}) with an exponential, obtaining in steady state
\begin{equation}\label{exponentialEq}
\addtolength{\belowdisplayskip}{-0.2cm}
\addtolength{\abovedisplayskip}{-0.2cm}
p_{i} = (1 -p_{i})\left( \eta + (1-\eta) \left[ 1- \exp \left(- \sum_{j} p_{j}A_{ij}\right) \right] \right)
\end{equation}
which we expand to second order using Eq.~(\ref{CtoFEq}) and $A u = \lambda u$,
\begin{equation}\label{cEq}
\addtolength{\belowdisplayskip}{-0.2cm}
\addtolength{\abovedisplayskip}{-0.2cm}
Cu_{i}+\epsilon_{i} = (A \epsilon)_{i} + \eta (1 - C u_{i}) + (1- \eta) \lambda C u_{i} - \left( \lambda + \frac{1}{2} \lambda^{2} \right) C^{2} u_{i}^{2} .
\end{equation}
To eliminate the error term $\epsilon_{i}$ from Eq.~(\ref{cEq}), we multiply by $v_{i}$, the $i$th entry of the left eigenvector corresponding to $\lambda$, and sum over $i$. We use the fact that $v^{T} A \epsilon = \lambda v^{T} \epsilon$, where $v^{T}$ denotes the transpose of $v$, and neglect the resulting small term $(1-\lambda) \sum_{i} v_{i} \epsilon_{i}$ close to the critical value $\lambda = 1$, obtaining
\begin{equation}\label{projectedOrderTwoEq}
\addtolength{\belowdisplayskip}{-0.2cm}
\addtolength{\abovedisplayskip}{-0.2cm}
C \langle u v \rangle = \eta ( \langle v \rangle - C \langle u v \rangle ) + (1 - \eta ) C \lambda \langle u v \rangle - \left( \lambda+ \frac{1}{2} \lambda^2\right) C^2 \langle v u^{2} \rangle.
\end{equation}
This equation is quadratic in $C$ [and therefore in $F$, via Eq.~(\ref{CtoFEq})] and linear in $\eta$, and may be easily solved for either. For $\eta = 0$ the nonzero solution for $F$ is
\begin{equation}\label{FzeroEq2}
\addtolength{\belowdisplayskip}{-0.2cm}
\addtolength{\abovedisplayskip}{-0.2cm}
F_{\eta = 0} = \frac{(\lambda-1)}{(\lambda + \frac{1}{2} \lambda^{2})} \frac{\langle uv \rangle \langle u \rangle}{\langle u^{2} v \rangle}.
\end{equation}
A more refined approximation than Eq.~(\ref{projectedOrderTwoEq}) can be obtained by repeating this process without expanding Eq.~(\ref{exponentialEq}), which yields the linear equation for $\eta$
\begin{equation}\label{exponentialSpectral}
\addtolength{\belowdisplayskip}{-0.4cm}
\addtolength{\abovedisplayskip}{-0.2cm}
C \langle u v \rangle = \sum_{i} ( 1 - C u_{i} ) ( \eta + (1 - \eta ) [ 1 - \exp ( - \lambda C u_{i} ) ] ).
\end{equation}

Before numerically testing our theory, we will explain how it
relates to previous results. For a network with correlations
between degrees at the ends of a randomly chosen edge (assortative
mixing by degree \cite{newmanAssortativity}), measured by the
correlation coefficient $\rho = \langle d_{i}^{in} d_{j}^{out}
\rangle_{e}/\langle d^{in} d^{out} \rangle$, with $\langle \cdot
\rangle_{e}$ denoting an average over edges, the largest
eigenvalue may be approximated by $\lambda \approx \rho \langle
d^{in} d^{out} \rangle / \langle d \rangle$
\cite{eigenvalueApproximation}. In the absence of assortativity,
when $\rho = 1$, $\lambda \approx \langle d^{in} d^{out}
\rangle/\langle d \rangle$. If, in addition, there are no correlations between
$d_{in}$ and $d_{out}$ (\textit{node degree correlations}) or if
the degree distribution is sufficiently homogenous, then $\langle
d^{in} d^{out} \rangle \approx \langle d \rangle^{2}$ and the
approximation reduces to $\lambda \approx \langle d \rangle$. In
the case of  Ref. \cite{kinouchiCopelli}, $\lambda \approx \langle
d \rangle$ applies, and in the case of Refs.
\cite{copelliCampos,wu}, $\lambda \approx \langle d^{in} d^{out}
\rangle/\langle d \rangle$ applies.

We test our theoretical results via direct simulation of the
Kinouchi-Copelli model on six categories of directed networks
with $N=10,000$ nodes: (category 1) Random networks with no node
degree correlation between $d^{in}$ and $d^{out}$; (category 2)
Random networks with maximal degree correlation, $d^{in} =
d^{out}$; (category 3) Random networks with moderate correlation
between $d^{in}$ and $d^{out}$; (category 4) Networks with power
law degree distribution with power law exponents $\gamma \in
[2.0,6.0]$, with and without node degree correlations; (category
5) Networks constructed with $\langle d \rangle = 1$, and
assortativity coefficient $\rho$ varying in $[0.7, 1.3]$; (category 6) Networks with weights which depend on the degree of the node from which the edge originates, $A_{ij} = \alpha / d_{i}^{out}$. 

We created networks in multiple steps: first, we created binary
networks ($A_{ij} \in \{ 0,1 \}$) with target degree distributions
as described below; next, we assigned a weight to each link, drawn
from a uniform distribution between 0 and 1; finally, we
calculated $\lambda$ for the resulting network and multiplied $A$
by a constant to rescale the largest eigenvalue to the targeted
eigenvalue. 
This process was restarted from the first step for
every network used in categories 1-4, creating a structurally
different network for each simulation. The initial binary networks in categories 
1-3 were Erd\H{o}s-R\'{e}nyi random networks, constructed by
linking any pair of nodes with probability $p = 10/N$
\cite{ERGraph}. Maximal degree correlation resulted from creating
undirected binary networks and then forcing $A_{ij} = A_{ji}$ for
$i < j$ while assigning weights. Moderate degree correlation
resulted from making undirected binary networks but allowing
$A_{ij} \neq A_{ji}$ when weights were assigned. The algorithms
for constructing the initial binary networks of categories 4-6
placed links randomly between nodes with specified in- and
out-degrees via the configuration model \cite{configurationModel}.
For this model, we generated in- and out-degree sequences from a
power law distribution of desired exponent $\gamma$ by calculating
the expected integer number of nodes with each integer degree,
from minimum degree 10 to maximum degree 200.  In creating
category 5 networks, we initially created one scale free network
with power law exponent $\gamma=2.5$ and $\lambda=1$. Then, to
change the degree of assortativity, we modified this original
network by choosing two links at random and swapping them if the
resulting swap would change the assortativity in the direction
desired. This process was repeated until a desired value of $\rho$
was achieved. Importantly, this swapping makes it possible to
leave the degree distributions of the network unchanged, while
still changing the assortative or disassortative properties of the
network as in \cite{eigenvalueApproximation,newmanAssortativity}.
Therefore, by this method we may maintain exactly the same degree
distribution and mean degree, yet modify $\lambda$ by virtue of $\lambda \propto \rho$.

\begin{figure}[t]
\vspace{-10pt} \centering \epsfig{file =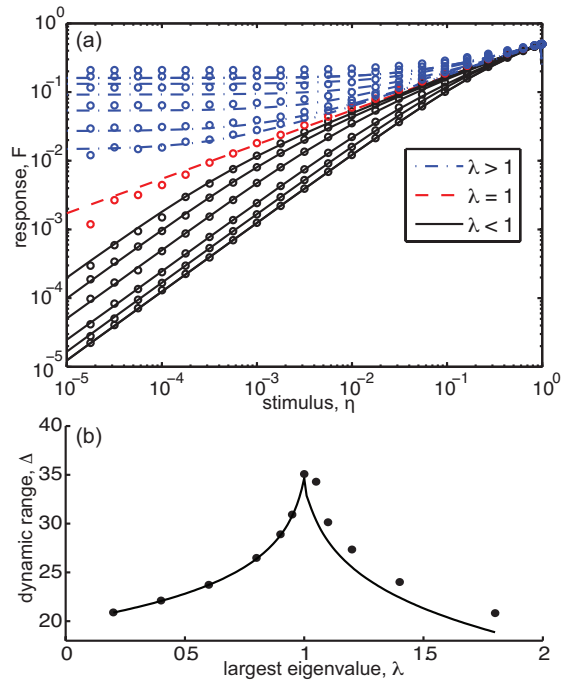, clip =
,width=0.95\linewidth } \caption{(color online) (a) Response $F$ vs. stimulus
$\eta$ for power law networks with exponent $\gamma=2.5$ and no
correlation between $d_{in}$ and $d_{out}$.
Eq.~(\ref{exponentialSpectral}) (lines) captures much of the
behavior of the simulation (circles), particularly for low levels
of $\eta$ and $F$, as expected from approximating
Eq.~(\ref{meanfieldEq}). (b) Dynamic range $\Delta$ is maximized
at $\lambda=1$ in both simulation results (circles) and
Eq.~(\ref{projectedOrderTwoEq}) (line).} \label{simulation43}
\vspace{-7pt}
\end{figure}

\begin{figure*}[t]
\vspace{-10pt} \centering \epsfig{file =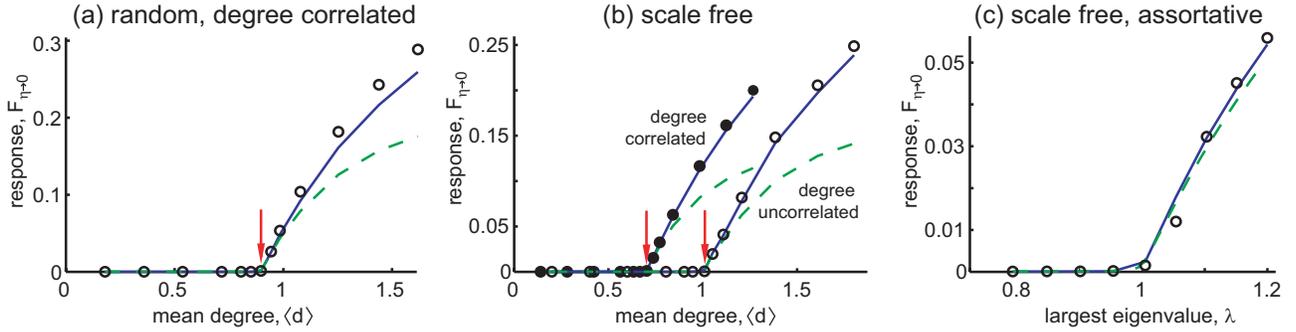, clip = ,width=0.95\linewidth }
\caption{ (color
online) $F_{\eta \to 0}$ obtained from direct numerical simulation
of the Kinouchi-Copelli model (symbols) plotted against $\langle d
\rangle$ (a, b) and $\lambda$ (c). Blue solid lines result from
iterating Eq.~(\ref{meanfieldEq}) and green dashed lines result
from Eq.~(\ref{FzeroEq2}). Small arrows show where $\lambda=1$
predicts a phase transition. (a) A set of random networks
(category 3) showing that criticality occurs at $\lambda = 1$
(arrow), but not $\langle d \rangle = 1$. (b) Criticality in scale
free networks (category 4) with node degree correlation also
occurs at $\lambda = 1$ (arrow), but not $\langle d \rangle = 1$.
(c) Category 5 networks are tuned through criticality by changing
assortativity, without changing the degree distributions and fixed
$\langle d \rangle = 1$.}\label{phaseTransitions}
\end{figure*}

In the six network types tested, results of simulations
unanimously confirm the hypothesis that criticality occurs only for
largest eigenvalue $\lambda=1$.  We present representative results
in Fig. \ref{simulation43} (a), noting that each line and set of
points corresponds to a single network realization, implying that
the effect of the largest eigenvalue on criticality is robust for
individual systems. Fig. \ref{simulation43} (a) shows the response
$F$ as a function of stimulus $\eta$ for scale-free networks  with
exponent $\gamma = 2.5$, constructed with no correlation between in-
and out-degree, highlighting the significant difference between
the regimes of $\lambda < 1$ and $\lambda >1$, with the critical
data corresponding to $\lambda=1$. The lines were obtained by
using Eqs.~(\ref{CtoFEq}) and (\ref{exponentialSpectral}). Fig.
\ref{simulation43} (b) shows $\Delta$ as a function of $\lambda$,
using $\eta_{high}=1$ and $\eta_{low}=0.01$, with the maximum
occurring at $\lambda = 1$. Similar results showing criticality and maximum dynamic range at
$\lambda = 1$ are obtained for networks of all categories 1-5.
Fig. \ref{phaseTransitions} shows $F_{\eta \to 0}$ for networks of
categories 3-5, confirming the transition predicted by the leading
order analysis in Eq.~(\ref{firstOrderEq}). The symbols show the
result of direct numerical simulation of the Kinouchi-Copelli
model, the solid lines were obtained by iterating
Eq.~(\ref{meanfieldEq}), and the dashed lines were obtained from
Eq.~(\ref{FzeroEq2}). Fig. \ref{phaseTransitions}(a) shows that
criticality occurs at $\lambda = 1$ (indicated by a vertical
arrow) rather than at $\langle d \rangle = 1$ for a category 3
random network. Fig. \ref{phaseTransitions}(b) shows that
criticality occurs at $\lambda = 1$ for scale-free networks
(category 4). Correlations between $d_{in}$ and $d_{out}$ affect
the point at which $\lambda = 1$ occurs (vertical arrows). In Fig.
\ref{phaseTransitions}(c), the mean degree was fixed at $\langle d
\rangle = 1$, while $\lambda$ was changed by modifying the
assortative coefficient $\rho$. As predicted by the theory, there
is a transition at $\lambda = 1$ even though the mean degree is
fixed.

We now explore the question of what network topology
will best enhance dynamic range.  In many of the systems we simulate, a majority of
the variation in dynamic range from one stimulus-response curve to another occurs
due to variation at the low stimulus end of the curve, since most
of the systems tend to saturate at around the same high stimulus
levels (though this may not be the case for neuronal network
experiments \cite{woody}). We therefore consider the following
approximate measure of dynamic range, $\Lambda$, obtained by
setting $\eta_{high}$ to one in the definition of $\Delta$,
$\Lambda = 10 \log_{10}{1/\eta_{*}}$, where $\eta_{*}$ is the
stimulus value corresponding to a lower threshold response
$F_{*}$. Since dynamic range is maximized
at criticality, we set $\lambda = 1$, solve
Eq.~(\ref{projectedOrderTwoEq}) for $\eta_{*}$, substitute it into
the definition of $\Lambda$ using Eq.~(\ref{CtoFEq}), retaining
the leading order behavior to get
\begin{equation}\label{momentEq}
\addtolength{\belowdisplayskip}{-0.2cm}
\addtolength{\abovedisplayskip}{-0.2cm}
\Lambda_{MAX} = 10 \log_{10} \frac{2}{3F_{*}^{2}} - 10 \log_{10} \frac{\langle v u^{2} \rangle}{\langle v \rangle \langle u \rangle ^{2}}.
\end{equation}

\begin{figure}[b]
\vspace{-5pt} \centering \epsfig{file =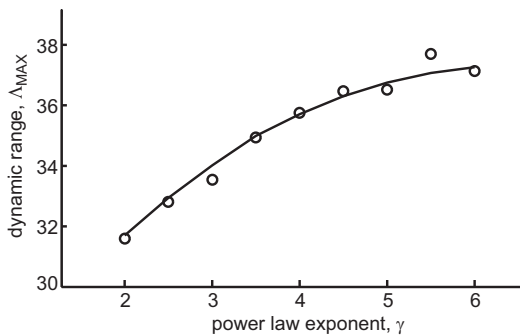, clip =
,width=0.8\linewidth } \caption{For power-law degree
distributions with $\lambda=1$, peak dynamic range increases
monotonically with network homogeneity, as measured by power law exponent $\gamma$. Simulations (circles) agree well with our predictions
[Eq.~(\ref{momentEq}); line]. } \label{maxDR}
\end{figure}

The first term of this equation shows that $\Lambda_{MAX}$ depends
on $F_{*}$. Since the entries of the right (left) dominant
eigenvector are a first order approximation to the in-degree
(out-degree) of the corresponding nodes \cite{degreeEigenvalue},
the second term suggests that maximum dynamic range should
increase (decrease) as the degree distribution becomes more
homogenous (heterogeneous). For example, consider the case of an
undirected, uncorrelated network, in which $v_{i}=u_{i} \approx
d_{i}$. The second term is then approximately $-10
\log_{10}{(\langle d^{3} \rangle / \langle d \rangle^{3})}$, which
is maximized when $d_{i}$ is independent of $i$. This corroborates
the numerical findings in Refs. \cite{kinouchiCopelli,wu} that
random graphs enhance dynamic range more than more
heterogeneous scale free graphs, and that the heterogeneity of the
degree distribution affects dynamic range \cite{wu}. To test our result,
we simulate scale free networks with different power law exponents
$\gamma \in [2.0,6.0]$, yet with $\lambda=1$ to maximize dynamic
range in each case. Results of simulation (circles) plotted against the
prediction of Eq.~(\ref{momentEq}) (line) are shown in Fig.
\ref{maxDR}.

In summary, we analytically predict and numerically confirm that
criticality and peak dynamic range occur in networks with largest
eigenvalue $\lambda = 1$.  This result holds for diverse network
topologies including random, scale-free, assortative, and/or
degree-correlated networks, and for networks in which edge weights are related to nodal degree, thus generalizing previous work.
Moreover, we find that homogeneous (heterogeneous) network
topologies result in higher (lower) dynamic range.  
Previous demonstrations of how $\lambda$ governs network dynamics in many other models (see \cite{degreeEigenvalue} and references therein) suggest that the generality of our findings may extend beyond the particular model studied here.
Previous model studies have shown that mutual information between stimulus and response is also maximized at criticality \cite{beggsPlenz}. 
Our findings suggest that peak mutual information will also be determined by $\lambda=1$, but verifying this will require additional investigation.
Taken together
with related experimental findings \cite{woody}, our results are
consistent with the hypotheses that 1) real brain networks operate
with $\lambda \approx 1$, and 2) if an organism benefits from
large dynamic range, then evolutionary pressures may act to
homogenize the network topology of the brain.

\begin{acknowledgments}
We thank Ed Ott and Dietmar Plenz for useful discussions. The work of Woodrow Shew was supported by the Intramural Research Program of the National Institute of Mental Health.
\end{acknowledgments}
\vspace{-0.2cm}

\begin{spacing}{0.90}
\bibliographystyle{plain}

\begin{thebibliography}{99}

\bibitem{dendriticTrees} L. L. Gollo \textit{et al.}, PLoS Comput. Biol {\bf 5(6)}: e10000402 (2009).
\bibitem{kinouchiCopelli} O. Kinouchi \textit{et al.}, Nature Physics {\bf 2}, 348 (2006).
\bibitem{epidemics} S. Gomez \textit{et al.}, EPL {\bf 89} 38009 (2010).
\bibitem{woody} W. L. Shew \textit{et al.} J. Neurosci {\bf 29(49)}:15595 (2009).
\bibitem{beggsPlenz} J. M. Beggs \textit{et al.}, J. Neurosci {\bf 23}: 11167-11177 (2003).
\bibitem{copelliCampos} M. Copelli \textit{et al.}, Eur Phys. J. B {\bf 56} 273 (2007).
\bibitem{wu} A. Wu \textit{et al.}, Phys. Rev. E {\bf 75} 032901 (2007).
\bibitem{foot} Our approach is easily generalized to include more refractory states. We also note that, in analogy to Ref.~\cite{ottPomerance}, our method can be generalized to include transmission delays and asynchronous updating. This will be discussed in a forthcoming publication.
\bibitem{locallyTreeLike} J. G. Restrepo \textit{et al.}, Phys. Rev. Lett {\bf 100}, 058701 (2008).
\bibitem{ottPomerance} E. Ott \textit{et al.}, Phys. Rev. E {\bf 79}, 056111 (2009).
\bibitem{genes} A. Pomerance \textit{et al.}, PNAS {\bf 106}, 20 (2009).
\bibitem{awakeMonkeys} T. Petermann \textit{et al.}, Proc. Natl. Acad. Sci. USA 106:15921Ð15926 (2009).
\bibitem{anesthetizedRats} E. D. Gireesh \textit{et al.}, Proc. Natl. Acad. Sci. USA 105:7576 Ð7581 (2008).
\bibitem{newmanAssortativity} M. E. J. Newman, Phys. Rev. E. {\bf 67}, 026126 (2003).
\bibitem{PFTheorem} C. R. MacCluer, SIAM Rev {\bf 42}:487 (2000).
\bibitem{robinsonGray} R.T. Gray \textit{et al.}, Neurocomputing {\bf 70}: 1000 (2007). 
\bibitem{rajanAbbott} K. Rajan \textit{et al.}, Phys. Rev. Lett {\bf 97}, 188104 (2006).
\bibitem{ERGraph} P. Erd\H{o}s \textit{et al.}, Publicationes Mathematicae {\bf 6} (1959).
\bibitem{configurationModel} M. E. J. Newman, SIAM Rev {\bf 45} 167 (2003).
\bibitem{eigenvalueApproximation} J. G. Restrepo \textit{et al.}, Phys. Rev. E {\bf 76}, 056119 (2007).
\bibitem{degreeEigenvalue} J. G. Restrepo \textit{et al.}, Phys. Rev. Lett. {\bf 97}, 094102 (2006).

\end{thebibliography}

\end{spacing}

\end{document}